\documentclass[reprint,aps,prb,superscriptaddress, nolongbibliography]{revtex4-2}

\usepackage{graphicx}% Include figure files
\usepackage{dcolumn}% Align table columns on decimal point
\usepackage{bm}% bold math
\usepackage{siunitx}
\usepackage{color}
\usepackage{booktabs}
\usepackage{miller} % for miller index of cristal axis

\definecolor{r}{rgb}{0.86,0.08,0.23}
\usepackage{hyperref}
\definecolor{r}{rgb}{0.86,0.08,0.23}

\definecolor{blue_n}{rgb}{0.,0.3,0.5}

\hypersetup{
backref=true, 
pagebackref=true,
hyperindex=true, 
colorlinks=true, 
breaklinks=true, 
citecolor = blue_n, 
urlcolor= blue_n,
linkcolor= black, 
pdftitle={Genuine and faux single G centers in carbon-implanted silicon}, 
pdfauthor={  Alrik~Durand, Yoann~Baron, F\'elix~Cache, Tobias~Herzig, Mario~Khoury, S\'ebastien~Pezzagna, Jan~Meijer, Marco~Abbarchi, Jean-Michel Hartmann, Shay Reboh, Marco~Abbarchi, Isabelle~Robert-Philip, Jean-Michel~G\'erard, Vincent~Jacques, Guillaume~Cassabois and Ana\"is~Dr\'eau } 
}

\begin{document}

    \title{Genuine and faux single G centers in carbon-implanted silicon}

	\author{Alrik Durand}\thanks{These authors contributed equally to this work.}	 	\affiliation{Laboratoire Charles Coulomb, Universit\'e de Montpellier and CNRS, 34095 Montpellier, France} 
	\author{Yoann Baron}\thanks{These authors contributed equally to this work.}			\affiliation{Laboratoire Charles Coulomb, Universit\'e de Montpellier and CNRS, 34095 Montpellier, France} 
	\author{F\'elix Cache}		\affiliation{Laboratoire Charles Coulomb, Universit\'e de Montpellier and CNRS, 34095 Montpellier, France} 
	\author{Tobias~Herzig}		\affiliation{Division of Applied Quantum Systems, Felix-Bloch Institute for Solid-State Physics, University Leipzig, Linn\'eestra\ss e 5, 04103 Leipzig, Germany} 
	\author{Mario~Khoury}	 	\affiliation{CNRS, Aix-Marseille Universit\'e, Centrale Marseille, IM2NP, UMR 7334, Campus de St. J\'er\^ome, 13397 Marseille, France} 
	\author{S\'ebastien~Pezzagna}	\affiliation{Division of Applied Quantum Systems, Felix-Bloch Institute for Solid-State Physics, University Leipzig, Linn\'eestra\ss e 5, 04103 Leipzig, Germany} 
	\author{Jan~Meijer}		\affiliation{Division of Applied Quantum Systems, Felix-Bloch Institute for Solid-State Physics, University Leipzig, Linn\'eestra\ss e 5, 04103 Leipzig, Germany} 
	\author{Jean-Michel~Hartmann}  \affiliation{Univ. Grenoble Alpes, CEA, LETI,  F-38000 Grenoble, France} 
	\author{Shay~Reboh}	\affiliation{Univ. Grenoble Alpes, CEA, LETI,  F-38000 Grenoble, France} 
	\author{Marco~Abbarchi}	 \affiliation{CNRS, Aix-Marseille Universit\'e, Centrale Marseille, IM2NP, UMR 7334, Campus de St. J\'er\^ome, 13397 Marseille, France} 
		\affiliation{Solnil, 95 Rue de la R\'epublique, 13002 Marseille, France}
	\author{Isabelle~Robert-Philip}   \affiliation{Laboratoire Charles Coulomb, Universit\'e de Montpellier and CNRS, 34095 Montpellier, France} 
	\author{Jean-Michel~G\'erard}	\affiliation{Univ. Grenoble Alpes, CEA, Grenoble INP, IRIG, PHELIQS, 38000 Grenoble, France} 
	\author{Vincent Jacques}	 \affiliation{Laboratoire Charles Coulomb, Universit\'e de Montpellier and CNRS, 34095 Montpellier, France} 
	\author{Guillaume Cassabois}	 \affiliation{Laboratoire Charles Coulomb, Universit\'e de Montpellier and CNRS, 34095 Montpellier, France} 
		\affiliation{Institut Universitaire de France, 75231 Paris, France.}
	\author{Ana\"is Dr\'eau}	\email{anais.dreau@umontpellier.fr}	 \affiliation{Laboratoire Charles Coulomb, Universit\'e de Montpellier and CNRS, 34095 Montpellier, France} \email{anais.dreau@umontpellier.fr}

    \begin{abstract}
Among the wide variety of single fluorescent defects investigated in silicon, numerous studies have focused on color centers with a zero-phonon line around $1.28 \mu$m and identified to a common carbon-complex in silicon, namely the G center. 
However, inconsistent estimates regarding their quantum efficiency cast doubt on the correct identification of these individual emitters. 
Through a comparative analysis of their single-photon emission properties, we demonstrate that these single color centers are split in two distinct families of point defects. 
A first family consists of the genuine single G centers with a well-identified microscopic structure and whose photoluminescence has been investigated on ensemble measurements since the 60's. 
The remaining defects belong to a new color center, which we will refer to as G$^{\star}$ center, whose atomic configuration has yet to be determined. 
These results provide a safeguard against future defect misidentifications, which is crucial for further development of quantum technologies relying on G or G$^{\star}$ center quantum properties.\\ 

    \end{abstract}

    \maketitle

The recent observation of single color centers in silicon has opened a new exploration path for silicon-based quantum technologies \cite{redjem_single_2020, hollenbach_engineering_2020, durand_broad_2021, baron_detection_2022, higginbottom_optical_2022, baron_single_2022, hollenbach_wafer-scale_2022, komza_indistinguishable_2022, gritsch_purcell_2023,  prabhu_individually_2023, saggio_cavity-enhanced_2023, redjem_all-silicon_2023, islam_cavity-enhanced_2024, johnston_cavity-coupled_2023, lee_high-efficiency_2023}. 
In less than three years, advanced single-defect spectroscopy has fueled the detection of more than 10 families of individual defects in silicon, able to emit non-classical, antibunched radiation in the near infrared \cite{redjem_single_2020, hollenbach_engineering_2020, durand_broad_2021,  baron_detection_2022, higginbottom_optical_2022, baron_single_2022, hollenbach_wafer-scale_2022, gritsch_purcell_2023}. 
Surprisingly, some of these defects were not previously referenced in the extensive literature about spectroscopic measurements on defect ensembles in silicon, likely because of their relative scarcity \cite{durand_broad_2021}. 
In view of applications in the fields of single photon sources or spin-photon interfaces, it is highly desirable for a color center to possess a high radiative quantum efficiency $\eta$. 
This value is defined as the probability after excitation to relax through photon emission rather than through a non-radiative channel. 
In this context, seemingly contradictory reports have been published for the G center in carbon-implanted silicon \cite{redjem_single_2020, durand_broad_2021,komza_indistinguishable_2022, prabhu_individually_2023, saggio_cavity-enhanced_2023, redjem_all-silicon_2023}, a defect with a zero-phonon line (ZPL) close to 1.28 µm \cite{thonke_new_1981, davies_carbon-related_1983, davies_optical_1989, beaufils_optical_2018}. 
A remarkably high quantum efficiency $\eta > 0.5$ has been reported in \cite{redjem_single_2020}, while other investigations at the single defect level \cite{komza_indistinguishable_2022, saggio_cavity-enhanced_2023} or on ensembles of G centers in optical cavities \cite{lefaucher_cavity-enhanced_2023} point at a much smaller $\eta$ in the few percents range at most. 

    \begin{figure}[h!]
        \includegraphics[width=0.9\columnwidth]{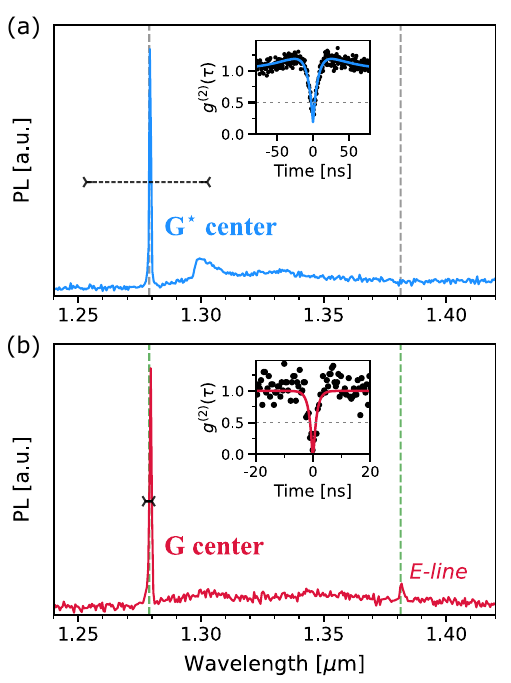}
        \caption{PL spectra measured for (a) a single G$^{\star}$  center and (b) a single G center in silicon, acquired at 11K and 30K respectively. 
        The vertical dashed lines indicate the wavelengths of the ZPL and E-line of the G center in silicon in PL reported in the literature \cite{thonke_new_1981, davies_carbon-related_1983, davies_optical_1989, beaufils_optical_2018}. 
        The dashed horizontal bars indicate the ZPL dispersion range on 41 and 39 defects, respectively. 
        Inset: autocorrelation function recorded on each defect, evidencing the single-photon emission with an antibunching at zero delay ${g^{(2)}(0) < 0.5}$ \cite{beveratos_room_2002}. 
        }
        \label{fig:spectra}
    \end{figure}

In this paper, we show that low-$\eta$ and high-$\eta$ color centers correspond to two different point defects, hereafter called G and G$^{\star}$ centers, respectively. 
Based on optical experiments at the single defect level, we establish the specific fingerprints of the G and G$^{\star}$ centers, and provide simple guidelines to identify them unambiguously among the wide variety of carbon-based color centers in silicon.

The experimental setup is a home-made low-temperature confocal microscope built up in a He closed-cycle cryostat (MyCryoFirm). 
Optical excitation of the sample is performed using a continuous laser at 532 nm and with a microscope objective (Olympus, LCPLN100XIR) mounted inside the cryostat vacuum chamber. 
The sample photoluminescence (PL) is collected by the same objective and measured by superconducting nanowire single-photon detectors with a detection efficiency of 78\% at 1.3 $\mu$m (SingleQuantum). 
Unless otherwise indicated, all measurements were conducted at 30K. 

Single defects are investigated in two different silicon-on-insulator (SOI) samples. 
The first sample (\#1) underwent a carbon-implantation over its entire surface at a fluence of $5 \times 10^{13}$ cm$^{-2}$, followed by a rapid thermal annealing during 20$\!$ s at 1,000$^{\circ}$C \cite{redjem_single_2020}.
The second sample (\#2) was locally implanted with carbon ions, then annealed with the same parameters as Sample \#1, and at last locally irradiated with protons \cite{baron_single_2022}. 
In Sample \#2, only the lowest doses, typically $\lesssim 1 \times 10^{11}$ cm$^{-2}$ for both carbon atoms and protons, lead to densities compatible with single color center isolation \cite{baron_single_2022}.  
Both SOI samples have a (001)-oriented top surface. 
The silicon top layer has a thickness of 220 nm for Sample~\#1 and 60 nm for Sample \#2.

We start by examining the photoluminescence (PL) spectra of single defects evidenced by antibunching g$^{(2)}(0) < 0.5$ (see Fig. \ref{fig:spectra} Insets).
The spectral emission of a single color center from Sample \#1 is shown in Figure \ref{fig:spectra}(a). 
It displays a strong ZPL at 1279 nm, corresponding to the 969.45 meV referenced ZPL energy of the G center in silicon \cite{davies_carbon-related_1983, beaufils_optical_2018}. 
A broad phonon replica with an energy of $\simeq 14.5$ meV is also found in the PL spectrum \cite{durand_broad_2021}.
This type of color center manifests a strong wavelength dispersion between defects: the ZPL position fluctuates in the range [1253-1303] nm (horizontal dashed line in Fig. \ref{fig:spectra}(a), for a set of 41 defects, with an average value of 1273 nm and a standard deviation of 12 nm. 
It should be noted that, although predominant, this family of fluorescent defects coexists with other unidentified single color centers in the Sample \#1 \cite{durand_broad_2021}.
On the contrary in Sample~\#2, 98\% of the fluorescent defects belong to the same family, connected to the typical spectrum displayed in Figure \ref{fig:spectra}(b) (the remaining 2\% have a broad spectrum with no detectable ZPL, similar to SD-6 defects in \cite{durand_broad_2021}).
This emission from an individual defect also shows an intense ZPL at the reference wavelength of the G center. 
However, here the zero-phonon line variations between centers are much smaller. Indeed, for 39 defects, the ZPL is only found between 1277 and 1280 nm, with a mean value of 1279 nm and a standard deviation of 0.5 nm.  
Furthermore, an extra line is visible in the PL spectrum at 1382 nm, hence $\simeq$ 72 meV lower than the ZPL energy. 
This emission line matches the E-line from the G center in silicon, associated to a local vibration mode (LVM) with a phonon energy of 71.9 meV reported in PL spectra on G-ensembles \cite{thonke_new_1981, davies_carbon-related_1983}. 
It can be seen on Figure \ref{fig:spectra}(a) that this E-line is not present in the PL spectrum of the individual defect from Sample \#1. 
 As a consequence, we can conclude that single color centers from Sample \#2 are genuine single G centers in silicon (Fig. \ref{fig:spectra}(b)). 
 Because of their markedly different spectral signatures, namely a strong ZPL dispersion and the lack of an E-line, single defects from Sample \#1 are not genuine G centers, and will be called G$^{\star}$ centers instead (Fig. \ref{fig:spectra}(a)).

    \begin{figure}
        \includegraphics[width=\columnwidth]{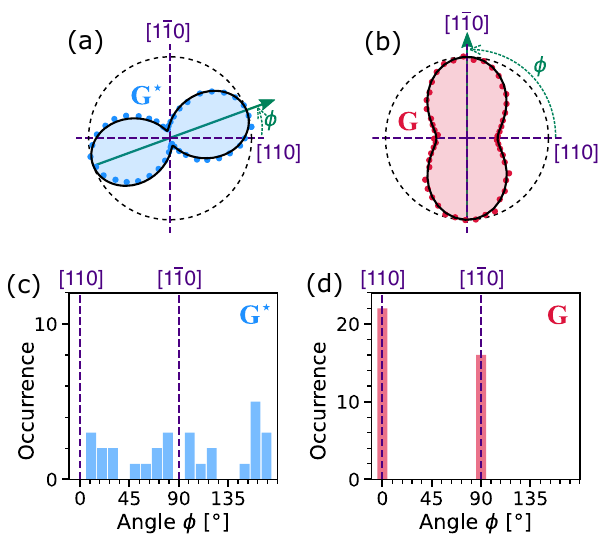}
        \caption{ Emission polarization diagrams recorded on (a) a single G$^{\star}$ center and (b) a single G center. 
        Solid lines represent data fitting using the function: $1-V +V\cos ^2 (\phi)$, where extracted visibilities are respectively $V^{\mathrm{(G^{\star})}} = 90\pm 2 \%$ and 
        $V^{\mathrm{(G)}} = 62 \pm 2 \%$. 
        Histograms of the diagram orientation angle $\phi$ for a set of (c) 30 single G$^{\mathrm{\star}}$ defects and (d) 38 individual G centers. }
        \label{fig:polar}
    \end{figure}

Another way to identify if a single emitter is a G or a G$^{\star}$ center is to analyze the polarization of its single photons. 
To this end, the emission signal of single centers is measured versus the angle of a polarizer installed before the PL collection fiber. 
As shown on Figure \ref{fig:polar}, two notable differences are seen on the emission polarization diagrams recorded on single G and G$^{\star}$ defects. 
First, the visibility of the G$^{\star}$ diagram is close to unity \cite{redjem_single_2020}, while the one of the G diagram is typically of $62 \pm 2 \%$ (Fig. \ref{fig:polar} (a,b)). 
Consequently the PL of G$^{\star}$ centers comes from a single emission dipole \cite{redjem_single_2020}, whereas the one from G centers is a result of the combination of several emission dipoles \cite{durand_notitle_2024}. 
Secondly, a statistical analysis performed on individual G defects reveals that the maximum intensity is always oriented along either $[110]$ or $[1\bar{1}0]$ crystal directions (Fig. \ref{fig:polar} (d)). 
These findings are in line with former spectroscopic reports from the 80's predicting an electric dipole along the $\langle 110\rangle$ axis \cite{foy_uniaxial_1981, thonke_new_1981}. 
On the contrary, a similar study carried out on single G$^{\star}$ centers shows that their emission dipole can point in a large number of directions, except crystal axes $[110]$ and $[1\bar{1}0]$ (Fig. \ref{fig:polar} (c)) \cite{redjem_single_2020}. 
The orientation distribution of the emission polarization diagrams of G and G$^{\star}$ defects are thus non-overlapping, allowing to directly associate the emitters to one or the other types of defect.

\begin{figure}
        \includegraphics[width=0.87\columnwidth]{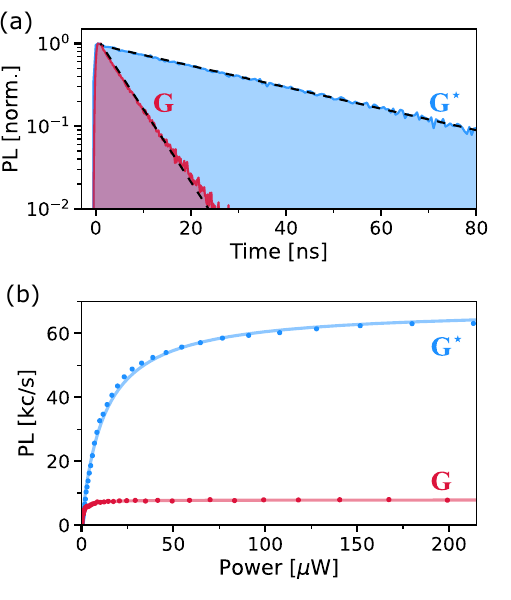}
        \caption{ (a) Excited-state lifetime measured on single G and G$^{\star}$ centers, under 150-ps pulsed laser excitation at 532 nm. 
        A fit by a mono-exponential function (dashed lines) gives the following decay times: $\tau^{\mathrm{(G)}} = 4.9 \pm 0.3$ ns and $\tau^{\mathrm{(G^{\star})}}= 33.4 \pm 0.5 $ns. 
        (b) Saturation curves recorded under continous pumping for individual G and G$^{\star}$ defects. 
        The solid lines show the data fitting with a standard saturation function $I_{sat}/(1+P_{sat}/P)$, leading to intensities at saturation  ${I_{sat} ^{\mathrm{(G)}}=7.9\pm0.1}\,$ kcounts.s$^{-1}$ and $I_{sat} ^{\mathrm{\mathrm{(G^{\star})}}} = 68\pm1$  kcounts.s$^{-1}$, and  to saturation powers ${P_{sat} ^{\mathrm{(G)}}=1.1~\pm0.1 }\, \mu$W and $P_{sat} ^{\mathrm{(G^{\star})}} = 12 \pm 1\,  \mu$W. 
        }
        \label{fig:dynamics}
    \end{figure}

Measuring the excited-state lifetime is an additional way to discriminate between G and G$^{\star}$ defects. 
Time-resolved measurements on single G centers show that they relax with a short timescale of roughly $\tau^{\mathrm{(G)}} = 4.9 \pm 0.3$ ns (Fig. \ref{fig:dynamics}(a)). 
This lifetime is very close to that reported on various ensembles of G centers in silicon \cite{beaufils_optical_2018, baron_single_2022}. 
On the contrary, single G$^{\star}$ centers exhibit a much longer lifetime, typically $\tau^{\mathrm{(G^{\star})}} \geq 30$  ns, as shown in Figure \ref{fig:dynamics}(b) (see also Ref. \cite{redjem_single_2020}).

Almost one order of magnitude is also observed in the single-photon count rates between the two types of color centers. 
Figure \ref{fig:dynamics} (b) displays the typical saturation curves recorded under continuous pumping for single G and G$^{\star}$ defects. 
While the G$^{\star}$ center count rate reaches $\simeq$ 65 kcounts.s$^{-1}$ for our experimental setup, the G center saturates around 7 kcounts.s$^{-1}$. 
In spite of their longer lifetime, single G$^{\star}$ defects are thus almost 10 times brighter at saturation than the genuine single G centers. 
We also note that the saturation powers  follow the same trend. 
Indeed, the signal of the G center starts to saturate for an excitation power $\simeq 1 \mu$W, while for the G$^{\star}$ defect, this occurs at $\simeq 12 \mu$W (Fig. \ref{fig:dynamics} (b)).

Comparing excited state lifetimes and photon count rates at saturation enables to assess the ratio of quantum efficiencies of individual G and G$^{\star}$ centers. 
By assuming that under continuous excitation at saturation all populations are prepared in the excited state for both defects, their respective quantum efficiencies, $\eta_{QE}^{\mathrm{(G)}} $ and $\eta_{QE}^{\mathrm{(G^{\star})}}$,  are related by the formula: 
\begin{equation}
\eta_{QE}^{\mathrm{(G)}} = \eta_{QE}^{\mathrm{(G^{\star})}} \frac{I_{sat}^{\mathrm{(G)}}} {I_{sat}^{\mathrm{(G^{\star})}}}
	\frac{\tau^{\mathrm{(G)}}}{\tau^{\mathrm{(G^{\star})}}}
	\frac{\eta_{coll}^{\mathrm{(G)}}}{\eta_{coll}^{\mathrm{(G^{\star})}}} \,, 
\end{equation}
where $I_{sat}^{\mathrm{(G)}}, I_{sat}^{\mathrm{(G^{\star})}}$ are intensities at saturation and $\eta_{coll}^{\mathrm{(G)}}$, $\eta_{coll}^{\mathrm{(G^{\star})}}$ collection efficiencies 
for single G and G$^{\star}$ centers, respectively (Fig. \ref{fig:dynamics}(b)). 
For a maximally collected dipole in the (001) plane, the collection efficiency varies with depth between 0.5\% and 2\% for the 220-nm silicon layer of Sample \#1 \cite{redjem_single_2020} [2.5\% and 4 \% for the 60-nm silicon layer of Sample \#2]. 
Considering that the brightest single defects have been selected in our experiments, the maximal value of the collection efficiency can be taken in both cases. 
Since the quantum efficiency of a single emitter cannot exceed 100$\%$ by definition, hence $\eta_{QE}^{\mathrm{(G^{\star})}} \leq 100 \%$, it follows immediately that the quantum efficiency of G centers is not greater than $1\%$. 
This low value is in agreement with previous studies \cite{lefaucher_cavity-enhanced_2023, komza_indistinguishable_2022} and explains the difficulty to observe a lifetime reduction induced by the Purcell effect for these emitters \cite{lefaucher_cavity-enhanced_2023,  saggio_cavity-enhanced_2023}. 
The quantum efficiency of single G$^{\star}$ centers is much greater. It has indeed been estimated to be at least $50\%$ \cite{redjem_single_2020}.

In conclusion, we have highlighted the existence of two families of single fluorescent defects in carbon-implanted silicon that emit with a ZPL around 1.28 $\mu$m. 
A first family consists of genuine single G centers in silicon, associated with the atomic configuration $\mathrm{C}_{(s)}$-$\mathrm{Si}_{(i)}$-$\mathrm{C}_{(s)}$ \cite{odonnell_origin_1983, song_bistable_1990, udvarhelyi_identification_2021}. 
Individual G centers are unambiguously identifiable by (i) an emission spectrum containing the E-line at 1382 nm, (ii) a short excited state lifetime of $\simeq$5 ns, (iii) a low quantum efficiency $\leq 1\%$ and (iv) a partial linear polarization but with well defined main polarization axis oriented either along $[110]$ or $[1\bar{1}0]$ crystal axes. 
On the other hand, individual G$^{\star}$ centers are characterized by (i) no E-line in their PL spectrum, (ii) an excited state lifetime typically above 30 ns, (iii) a bright emission connected to a high quantum efficiency $> 50\%$ and (iv) a single emission dipole whose orientation deviates from $[110]$ or $[1\bar{1}0]$. 
Following these criteria, the single emitters investigated in \cite{redjem_single_2020, hollenbach_engineering_2020, durand_broad_2021, redjem_all-silicon_2023} are likely G$^{\star}$ centers, whereas the genuine G centers are observed in Ref. \cite{baron_single_2022, hollenbach_wafer-scale_2022,  prabhu_individually_2023, komza_indistinguishable_2022, saggio_cavity-enhanced_2023}. 
Note that the fingerprints given here for G and G$^{\star}$ centers can vary depending on their environment. 
For instance, a thicker top silicon layer can reduce the inhomogeneous linewidth of G$^{\star}$ centers \cite{hollenbach_engineering_2020} compared to our findings, and cavity integration can favor one emission dipole of the G centers as in \cite{saggio_cavity-enhanced_2023}.

The G center has a head start since its microscopic structure has been identified \cite{odonnell_origin_1983, song_bistable_1990, udvarhelyi_identification_2021} and it can be selectively fabricated at single-defect scale through ion implantation \cite{baron_single_2022, hollenbach_wafer-scale_2022}. 
Moreover, it could host a spin qubit since a spin resonance linked to a metastable spin triplet has been detected optically on ensembles of G centers in silicon \cite{lee_optical_1982, odonnell_origin_1983}. 
In contrast, the atomic configuration of the G$^{\star}$ center is currently unknown, as well as its spin properties, and no selective fabrication method is known to date for such centers \cite{durand_broad_2021}. 
Nevertheless, its high quantum efficiency combined with single-dipole emission makes it a very promising color center to fabricate deterministic telecom single-photon sources integrated in silicon. 
Further investigations, including defect engineering to look for isotope shifts and \textit{ab initio} calculations, are required to elucidate the G$^{\star}$ center origin and its potential connection to the G center in silicon.

\section*{Acknowledgments}

We acknowledge funding from the French National Research Agency (ANR) through the projects OCTOPUS (No. ANR-18-CE47-0013-01) and QUASSIC (No. ANR-18-ERC2-0005-01), the Plan France 2030 through the project OQuLuS ANR-22-PETQ-0013, the Occitanie region through the SITEQ contract and the European Research Council (ERC) under the European Union’s Horizon 2020 research and innovation programme (project SILEQS, Grant No. 101042075). 
A. Durand acknowledges support from the French DGA.

\bibliography{bib_genuine_faux_G}

\begin{thebibliography}{27}%
\makeatletter
\providecommand \@ifxundefined [1]{%
 \@ifx{#1\undefined}
}%
\providecommand \@ifnum [1]{%
 \ifnum #1\expandafter \@firstoftwo
 \else \expandafter \@secondoftwo
 \fi
}%
\providecommand \@ifx [1]{%
 \ifx #1\expandafter \@firstoftwo
 \else \expandafter \@secondoftwo
 \fi
}%
\providecommand \natexlab [1]{#1}%
\providecommand \enquote  [1]{``#1''}%
\providecommand \bibnamefont  [1]{#1}%
\providecommand \bibfnamefont [1]{#1}%
\providecommand \citenamefont [1]{#1}%
\providecommand \href@noop [0]{\@secondoftwo}%
\providecommand \href [0]{\begingroup \@sanitize@url \@href}%
\providecommand \@href[1]{\@@startlink{#1}\@@href}%
\providecommand \@@href[1]{\endgroup#1\@@endlink}%
\providecommand \@sanitize@url [0]{\catcode `\\12\catcode `\$12\catcode
  `\&12\catcode `\#12\catcode `\^12\catcode `\_12\catcode `\%12\relax}%
\providecommand \@@startlink[1]{}%
\providecommand \@@endlink[0]{}%
\providecommand \url  [0]{\begingroup\@sanitize@url \@url }%
\providecommand \@url [1]{\endgroup\@href {#1}{\urlprefix }}%
\providecommand \urlprefix  [0]{URL }%
\providecommand \Eprint [0]{\href }%
\providecommand \doibase [0]{https://doi.org/}%
\providecommand \selectlanguage [0]{\@gobble}%
\providecommand \bibinfo  [0]{\@secondoftwo}%
\providecommand \bibfield  [0]{\@secondoftwo}%
\providecommand \translation [1]{[#1]}%
\providecommand \BibitemOpen [0]{}%
\providecommand \bibitemStop [0]{}%
\providecommand \bibitemNoStop [0]{.\EOS\space}%
\providecommand \EOS [0]{\spacefactor3000\relax}%
\providecommand \BibitemShut  [1]{\csname bibitem#1\endcsname}%
\let\auto@bib@innerbib\@empty
%</preamble>
\bibitem [{\citenamefont {Redjem}\ \emph {et~al.}(2020)\citenamefont {Redjem},
  \citenamefont {Durand}, \citenamefont {Herzig}, \citenamefont {Benali},
  \citenamefont {Pezzagna}, \citenamefont {Meijer}, \citenamefont {Kuznetsov},
  \citenamefont {Nguyen}, \citenamefont {Cueff}, \citenamefont {Gérard},
  \citenamefont {Robert-Philip}, \citenamefont {Gil}, \citenamefont {Caliste},
  \citenamefont {Pochet}, \citenamefont {Abbarchi}, \citenamefont {Jacques},
  \citenamefont {Dréau},\ and\ \citenamefont
  {Cassabois}}]{redjem_single_2020}%
  \BibitemOpen
  \bibfield  {author} {\bibinfo {author} {\bibfnamefont {W.}~\bibnamefont
  {Redjem}}, \bibinfo {author} {\bibfnamefont {A.}~\bibnamefont {Durand}},
  \bibinfo {author} {\bibfnamefont {T.}~\bibnamefont {Herzig}}, \bibinfo
  {author} {\bibfnamefont {A.}~\bibnamefont {Benali}}, \bibinfo {author}
  {\bibfnamefont {S.}~\bibnamefont {Pezzagna}}, \bibinfo {author}
  {\bibfnamefont {J.}~\bibnamefont {Meijer}}, \bibinfo {author} {\bibfnamefont
  {A.~Y.}\ \bibnamefont {Kuznetsov}}, \bibinfo {author} {\bibfnamefont {H.~S.}\
  \bibnamefont {Nguyen}}, \bibinfo {author} {\bibfnamefont {S.}~\bibnamefont
  {Cueff}}, \bibinfo {author} {\bibfnamefont {J.-M.}\ \bibnamefont {Gérard}},
  \bibinfo {author} {\bibfnamefont {I.}~\bibnamefont {Robert-Philip}}, \bibinfo
  {author} {\bibfnamefont {B.}~\bibnamefont {Gil}}, \bibinfo {author}
  {\bibfnamefont {D.}~\bibnamefont {Caliste}}, \bibinfo {author} {\bibfnamefont
  {P.}~\bibnamefont {Pochet}}, \bibinfo {author} {\bibfnamefont
  {M.}~\bibnamefont {Abbarchi}}, \bibinfo {author} {\bibfnamefont
  {V.}~\bibnamefont {Jacques}}, \bibinfo {author} {\bibfnamefont
  {A.}~\bibnamefont {Dréau}},\ and\ \bibinfo {author} {\bibfnamefont
  {G.}~\bibnamefont {Cassabois}},\ }\href
  {https://doi.org/10.1038/s41928-020-00499-0} {\bibfield  {journal} {\bibinfo
  {journal} {Nature Electronics}\ }\textbf {\bibinfo {volume} {3}},\ \bibinfo
  {pages} {738} (\bibinfo {year} {2020})}\BibitemShut {NoStop}%
\bibitem [{\citenamefont {Hollenbach}\ \emph {et~al.}(2020)\citenamefont
  {Hollenbach}, \citenamefont {Berencén}, \citenamefont {Kentsch},
  \citenamefont {Helm},\ and\ \citenamefont
  {Astakhov}}]{hollenbach_engineering_2020}%
  \BibitemOpen
  \bibfield  {author} {\bibinfo {author} {\bibfnamefont {M.}~\bibnamefont
  {Hollenbach}}, \bibinfo {author} {\bibfnamefont {Y.}~\bibnamefont
  {Berencén}}, \bibinfo {author} {\bibfnamefont {U.}~\bibnamefont {Kentsch}},
  \bibinfo {author} {\bibfnamefont {M.}~\bibnamefont {Helm}},\ and\ \bibinfo
  {author} {\bibfnamefont {G.~V.}\ \bibnamefont {Astakhov}},\ }\href
  {https://doi.org/10.1364/OE.397377} {\bibfield  {journal} {\bibinfo
  {journal} {Optics Express}\ }\textbf {\bibinfo {volume} {28}},\ \bibinfo
  {pages} {26111} (\bibinfo {year} {2020})}\BibitemShut {NoStop}%
\bibitem [{\citenamefont {Durand}\ \emph {et~al.}(2021)\citenamefont {Durand},
  \citenamefont {Baron}, \citenamefont {Redjem}, \citenamefont {Herzig},
  \citenamefont {Benali}, \citenamefont {Pezzagna}, \citenamefont {Meijer},
  \citenamefont {Kuznetsov}, \citenamefont {Gérard}, \citenamefont
  {Robert-Philip}, \citenamefont {Abbarchi}, \citenamefont {Jacques},
  \citenamefont {Cassabois},\ and\ \citenamefont {Dréau}}]{durand_broad_2021}%
  \BibitemOpen
  \bibfield  {author} {\bibinfo {author} {\bibfnamefont {A.}~\bibnamefont
  {Durand}}, \bibinfo {author} {\bibfnamefont {Y.}~\bibnamefont {Baron}},
  \bibinfo {author} {\bibfnamefont {W.}~\bibnamefont {Redjem}}, \bibinfo
  {author} {\bibfnamefont {T.}~\bibnamefont {Herzig}}, \bibinfo {author}
  {\bibfnamefont {A.}~\bibnamefont {Benali}}, \bibinfo {author} {\bibfnamefont
  {S.}~\bibnamefont {Pezzagna}}, \bibinfo {author} {\bibfnamefont
  {J.}~\bibnamefont {Meijer}}, \bibinfo {author} {\bibfnamefont
  {A.}~\bibnamefont {Kuznetsov}}, \bibinfo {author} {\bibfnamefont {J.-M.}\
  \bibnamefont {Gérard}}, \bibinfo {author} {\bibfnamefont {I.}~\bibnamefont
  {Robert-Philip}}, \bibinfo {author} {\bibfnamefont {M.}~\bibnamefont
  {Abbarchi}}, \bibinfo {author} {\bibfnamefont {V.}~\bibnamefont {Jacques}},
  \bibinfo {author} {\bibfnamefont {G.}~\bibnamefont {Cassabois}},\ and\
  \bibinfo {author} {\bibfnamefont {A.}~\bibnamefont {Dréau}},\ }\href
  {https://doi.org/10.1103/PhysRevLett.126.083602} {\bibfield  {journal}
  {\bibinfo  {journal} {Physical Review Letters}\ }\textbf {\bibinfo {volume}
  {126}},\ \bibinfo {pages} {083602} (\bibinfo {year} {2021})}\BibitemShut
  {NoStop}%
\bibitem [{\citenamefont {Baron}\ \emph
  {et~al.}(2022{\natexlab{a}})\citenamefont {Baron}, \citenamefont {Durand},
  \citenamefont {Udvarhelyi}, \citenamefont {Herzig}, \citenamefont {Khoury},
  \citenamefont {Pezzagna}, \citenamefont {Meijer}, \citenamefont
  {Robert-Philip}, \citenamefont {Abbarchi}, \citenamefont {Hartmann},
  \citenamefont {Mazzocchi}, \citenamefont {Gérard}, \citenamefont {Gali},
  \citenamefont {Jacques}, \citenamefont {Cassabois},\ and\ \citenamefont
  {Dréau}}]{baron_detection_2022}%
  \BibitemOpen
  \bibfield  {author} {\bibinfo {author} {\bibfnamefont {Y.}~\bibnamefont
  {Baron}}, \bibinfo {author} {\bibfnamefont {A.}~\bibnamefont {Durand}},
  \bibinfo {author} {\bibfnamefont {P.}~\bibnamefont {Udvarhelyi}}, \bibinfo
  {author} {\bibfnamefont {T.}~\bibnamefont {Herzig}}, \bibinfo {author}
  {\bibfnamefont {M.}~\bibnamefont {Khoury}}, \bibinfo {author} {\bibfnamefont
  {S.}~\bibnamefont {Pezzagna}}, \bibinfo {author} {\bibfnamefont
  {J.}~\bibnamefont {Meijer}}, \bibinfo {author} {\bibfnamefont
  {I.}~\bibnamefont {Robert-Philip}}, \bibinfo {author} {\bibfnamefont
  {M.}~\bibnamefont {Abbarchi}}, \bibinfo {author} {\bibfnamefont {J.-M.}\
  \bibnamefont {Hartmann}}, \bibinfo {author} {\bibfnamefont {V.}~\bibnamefont
  {Mazzocchi}}, \bibinfo {author} {\bibfnamefont {J.-M.}\ \bibnamefont
  {Gérard}}, \bibinfo {author} {\bibfnamefont {A.}~\bibnamefont {Gali}},
  \bibinfo {author} {\bibfnamefont {V.}~\bibnamefont {Jacques}}, \bibinfo
  {author} {\bibfnamefont {G.}~\bibnamefont {Cassabois}},\ and\ \bibinfo
  {author} {\bibfnamefont {A.}~\bibnamefont {Dréau}},\ }\href
  {https://doi.org/10.1021/acsphotonics.2c00336} {\bibfield  {journal}
  {\bibinfo  {journal} {ACS Photonics}\ }\textbf {\bibinfo {volume} {9}},\
  \bibinfo {pages} {2337} (\bibinfo {year} {2022}{\natexlab{a}})}\BibitemShut
  {NoStop}%
\bibitem [{\citenamefont {Higginbottom}\ \emph {et~al.}(2022)\citenamefont
  {Higginbottom}, \citenamefont {Kurkjian}, \citenamefont {Chartrand},
  \citenamefont {Kazemi}, \citenamefont {Brunelle}, \citenamefont {MacQuarrie},
  \citenamefont {Klein}, \citenamefont {Lee-Hone}, \citenamefont {Stacho},
  \citenamefont {Ruether}, \citenamefont {Bowness}, \citenamefont {Bergeron},
  \citenamefont {DeAbreu}, \citenamefont {Harrigan}, \citenamefont
  {Kanaganayagam}, \citenamefont {Marsden}, \citenamefont {Richards},
  \citenamefont {Stott}, \citenamefont {Roorda}, \citenamefont {Morse},
  \citenamefont {Thewalt},\ and\ \citenamefont
  {Simmons}}]{higginbottom_optical_2022}%
  \BibitemOpen
  \bibfield  {author} {\bibinfo {author} {\bibfnamefont {D.~B.}\ \bibnamefont
  {Higginbottom}}, \bibinfo {author} {\bibfnamefont {A.~T.~K.}\ \bibnamefont
  {Kurkjian}}, \bibinfo {author} {\bibfnamefont {C.}~\bibnamefont {Chartrand}},
  \bibinfo {author} {\bibfnamefont {M.}~\bibnamefont {Kazemi}}, \bibinfo
  {author} {\bibfnamefont {N.~A.}\ \bibnamefont {Brunelle}}, \bibinfo {author}
  {\bibfnamefont {E.~R.}\ \bibnamefont {MacQuarrie}}, \bibinfo {author}
  {\bibfnamefont {J.~R.}\ \bibnamefont {Klein}}, \bibinfo {author}
  {\bibfnamefont {N.~R.}\ \bibnamefont {Lee-Hone}}, \bibinfo {author}
  {\bibfnamefont {J.}~\bibnamefont {Stacho}}, \bibinfo {author} {\bibfnamefont
  {M.}~\bibnamefont {Ruether}}, \bibinfo {author} {\bibfnamefont
  {C.}~\bibnamefont {Bowness}}, \bibinfo {author} {\bibfnamefont
  {L.}~\bibnamefont {Bergeron}}, \bibinfo {author} {\bibfnamefont
  {A.}~\bibnamefont {DeAbreu}}, \bibinfo {author} {\bibfnamefont {S.~R.}\
  \bibnamefont {Harrigan}}, \bibinfo {author} {\bibfnamefont {J.}~\bibnamefont
  {Kanaganayagam}}, \bibinfo {author} {\bibfnamefont {D.~W.}\ \bibnamefont
  {Marsden}}, \bibinfo {author} {\bibfnamefont {T.~S.}\ \bibnamefont
  {Richards}}, \bibinfo {author} {\bibfnamefont {L.~A.}\ \bibnamefont {Stott}},
  \bibinfo {author} {\bibfnamefont {S.}~\bibnamefont {Roorda}}, \bibinfo
  {author} {\bibfnamefont {K.~J.}\ \bibnamefont {Morse}}, \bibinfo {author}
  {\bibfnamefont {M.~L.~W.}\ \bibnamefont {Thewalt}},\ and\ \bibinfo {author}
  {\bibfnamefont {S.}~\bibnamefont {Simmons}},\ }\href
  {https://doi.org/10.1038/s41586-022-04821-y} {\bibfield  {journal} {\bibinfo
  {journal} {Nature}\ }\textbf {\bibinfo {volume} {607}},\ \bibinfo {pages}
  {266} (\bibinfo {year} {2022})}\BibitemShut {NoStop}%
\bibitem [{\citenamefont {Baron}\ \emph
  {et~al.}(2022{\natexlab{b}})\citenamefont {Baron}, \citenamefont {Durand},
  \citenamefont {Herzig}, \citenamefont {Khoury}, \citenamefont {Pezzagna},
  \citenamefont {Meijer}, \citenamefont {Robert-Philip}, \citenamefont
  {Abbarchi}, \citenamefont {Hartmann}, \citenamefont {Reboh}, \citenamefont
  {Gérard}, \citenamefont {Jacques}, \citenamefont {Cassabois},\ and\
  \citenamefont {Dréau}}]{baron_single_2022}%
  \BibitemOpen
  \bibfield  {author} {\bibinfo {author} {\bibfnamefont {Y.}~\bibnamefont
  {Baron}}, \bibinfo {author} {\bibfnamefont {A.}~\bibnamefont {Durand}},
  \bibinfo {author} {\bibfnamefont {T.}~\bibnamefont {Herzig}}, \bibinfo
  {author} {\bibfnamefont {M.}~\bibnamefont {Khoury}}, \bibinfo {author}
  {\bibfnamefont {S.}~\bibnamefont {Pezzagna}}, \bibinfo {author}
  {\bibfnamefont {J.}~\bibnamefont {Meijer}}, \bibinfo {author} {\bibfnamefont
  {I.}~\bibnamefont {Robert-Philip}}, \bibinfo {author} {\bibfnamefont
  {M.}~\bibnamefont {Abbarchi}}, \bibinfo {author} {\bibfnamefont {J.-M.}\
  \bibnamefont {Hartmann}}, \bibinfo {author} {\bibfnamefont {S.}~\bibnamefont
  {Reboh}}, \bibinfo {author} {\bibfnamefont {J.-M.}\ \bibnamefont {Gérard}},
  \bibinfo {author} {\bibfnamefont {V.}~\bibnamefont {Jacques}}, \bibinfo
  {author} {\bibfnamefont {G.}~\bibnamefont {Cassabois}},\ and\ \bibinfo
  {author} {\bibfnamefont {A.}~\bibnamefont {Dréau}},\ }\href
  {https://doi.org/10.1063/5.0097407} {\bibfield  {journal} {\bibinfo
  {journal} {Applied Physics Letters}\ }\textbf {\bibinfo {volume} {121}},\
  \bibinfo {pages} {084003} (\bibinfo {year} {2022}{\natexlab{b}})}\BibitemShut
  {NoStop}%
\bibitem [{\citenamefont {Hollenbach}\ \emph {et~al.}(2022)\citenamefont
  {Hollenbach}, \citenamefont {Klingner}, \citenamefont {Jagtap}, \citenamefont
  {Bischoff}, \citenamefont {Fowley}, \citenamefont {Kentsch}, \citenamefont
  {Hlawacek}, \citenamefont {Erbe}, \citenamefont {Abrosimov}, \citenamefont
  {Helm}, \citenamefont {Berencén},\ and\ \citenamefont
  {Astakhov}}]{hollenbach_wafer-scale_2022}%
  \BibitemOpen
  \bibfield  {author} {\bibinfo {author} {\bibfnamefont {M.}~\bibnamefont
  {Hollenbach}}, \bibinfo {author} {\bibfnamefont {N.}~\bibnamefont
  {Klingner}}, \bibinfo {author} {\bibfnamefont {N.~S.}\ \bibnamefont
  {Jagtap}}, \bibinfo {author} {\bibfnamefont {L.}~\bibnamefont {Bischoff}},
  \bibinfo {author} {\bibfnamefont {C.}~\bibnamefont {Fowley}}, \bibinfo
  {author} {\bibfnamefont {U.}~\bibnamefont {Kentsch}}, \bibinfo {author}
  {\bibfnamefont {G.}~\bibnamefont {Hlawacek}}, \bibinfo {author}
  {\bibfnamefont {A.}~\bibnamefont {Erbe}}, \bibinfo {author} {\bibfnamefont
  {N.~V.}\ \bibnamefont {Abrosimov}}, \bibinfo {author} {\bibfnamefont
  {M.}~\bibnamefont {Helm}}, \bibinfo {author} {\bibfnamefont {Y.}~\bibnamefont
  {Berencén}},\ and\ \bibinfo {author} {\bibfnamefont {G.~V.}\ \bibnamefont
  {Astakhov}},\ }\href {https://doi.org/10.1038/s41467-022-35051-5} {\bibfield
  {journal} {\bibinfo  {journal} {Nature Communications}\ }\textbf {\bibinfo
  {volume} {13}},\ \bibinfo {pages} {7683} (\bibinfo {year}
  {2022})}\BibitemShut {NoStop}%
\bibitem [{\citenamefont {Komza}\ \emph {et~al.}(2022)\citenamefont {Komza},
  \citenamefont {Samutpraphoot}, \citenamefont {Odeh}, \citenamefont {Tang},
  \citenamefont {Mathew}, \citenamefont {Chang}, \citenamefont {Song},
  \citenamefont {Kim}, \citenamefont {Xiong}, \citenamefont {Hautier},\ and\
  \citenamefont {Sipahigil}}]{komza_indistinguishable_2022}%
  \BibitemOpen
  \bibfield  {author} {\bibinfo {author} {\bibfnamefont {L.}~\bibnamefont
  {Komza}}, \bibinfo {author} {\bibfnamefont {P.}~\bibnamefont
  {Samutpraphoot}}, \bibinfo {author} {\bibfnamefont {M.}~\bibnamefont {Odeh}},
  \bibinfo {author} {\bibfnamefont {Y.-L.}\ \bibnamefont {Tang}}, \bibinfo
  {author} {\bibfnamefont {M.}~\bibnamefont {Mathew}}, \bibinfo {author}
  {\bibfnamefont {J.}~\bibnamefont {Chang}}, \bibinfo {author} {\bibfnamefont
  {H.}~\bibnamefont {Song}}, \bibinfo {author} {\bibfnamefont {M.-K.}\
  \bibnamefont {Kim}}, \bibinfo {author} {\bibfnamefont {Y.}~\bibnamefont
  {Xiong}}, \bibinfo {author} {\bibfnamefont {G.}~\bibnamefont {Hautier}},\
  and\ \bibinfo {author} {\bibfnamefont {A.}~\bibnamefont {Sipahigil}},\ }\href
  {https://doi.org/10.48550/arXiv.2211.09305} {\bibinfo {title}
  {Indistinguishable photons from an artificial atom in silicon photonics}}
  (\bibinfo {year} {2022}),\ \bibinfo {note} {arXiv:2211.09305 [cond-mat,
  physics:physics, physics:quant-ph]}\BibitemShut {NoStop}%
\bibitem [{\citenamefont {Gritsch}\ \emph {et~al.}(2023)\citenamefont
  {Gritsch}, \citenamefont {Ulanowski},\ and\ \citenamefont
  {Reiserer}}]{gritsch_purcell_2023}%
  \BibitemOpen
  \bibfield  {author} {\bibinfo {author} {\bibfnamefont {A.}~\bibnamefont
  {Gritsch}}, \bibinfo {author} {\bibfnamefont {A.}~\bibnamefont {Ulanowski}},\
  and\ \bibinfo {author} {\bibfnamefont {A.}~\bibnamefont {Reiserer}},\ }\href
  {https://doi.org/10.1364/OPTICA.486167} {\bibfield  {journal} {\bibinfo
  {journal} {Optica}\ }\textbf {\bibinfo {volume} {10}},\ \bibinfo {pages}
  {783} (\bibinfo {year} {2023})}\BibitemShut {NoStop}%
\bibitem [{\citenamefont {Prabhu}\ \emph {et~al.}(2023)\citenamefont {Prabhu},
  \citenamefont {Errando-Herranz}, \citenamefont {De~Santis}, \citenamefont
  {Christen}, \citenamefont {Chen}, \citenamefont {Gerlach},\ and\
  \citenamefont {Englund}}]{prabhu_individually_2023}%
  \BibitemOpen
  \bibfield  {author} {\bibinfo {author} {\bibfnamefont {M.}~\bibnamefont
  {Prabhu}}, \bibinfo {author} {\bibfnamefont {C.}~\bibnamefont
  {Errando-Herranz}}, \bibinfo {author} {\bibfnamefont {L.}~\bibnamefont
  {De~Santis}}, \bibinfo {author} {\bibfnamefont {I.}~\bibnamefont {Christen}},
  \bibinfo {author} {\bibfnamefont {C.}~\bibnamefont {Chen}}, \bibinfo {author}
  {\bibfnamefont {C.}~\bibnamefont {Gerlach}},\ and\ \bibinfo {author}
  {\bibfnamefont {D.}~\bibnamefont {Englund}},\ }\href
  {https://doi.org/10.1038/s41467-023-37655-x} {\bibfield  {journal} {\bibinfo
  {journal} {Nature Communications}\ }\textbf {\bibinfo {volume} {14}},\
  \bibinfo {pages} {2380} (\bibinfo {year} {2023})}\BibitemShut {NoStop}%
\bibitem [{\citenamefont {Saggio}\ \emph {et~al.}(2023)\citenamefont {Saggio},
  \citenamefont {Errando-Herranz}, \citenamefont {Gyger}, \citenamefont
  {Panuski}, \citenamefont {Prabhu}, \citenamefont {De~Santis}, \citenamefont
  {Christen}, \citenamefont {Ornelas-Huerta}, \citenamefont {Raniwala},
  \citenamefont {Gerlach}, \citenamefont {Colangelo},\ and\ \citenamefont
  {Englund}}]{saggio_cavity-enhanced_2023}%
  \BibitemOpen
  \bibfield  {author} {\bibinfo {author} {\bibfnamefont {V.}~\bibnamefont
  {Saggio}}, \bibinfo {author} {\bibfnamefont {C.}~\bibnamefont
  {Errando-Herranz}}, \bibinfo {author} {\bibfnamefont {S.}~\bibnamefont
  {Gyger}}, \bibinfo {author} {\bibfnamefont {C.}~\bibnamefont {Panuski}},
  \bibinfo {author} {\bibfnamefont {M.}~\bibnamefont {Prabhu}}, \bibinfo
  {author} {\bibfnamefont {L.}~\bibnamefont {De~Santis}}, \bibinfo {author}
  {\bibfnamefont {I.}~\bibnamefont {Christen}}, \bibinfo {author}
  {\bibfnamefont {D.}~\bibnamefont {Ornelas-Huerta}}, \bibinfo {author}
  {\bibfnamefont {H.}~\bibnamefont {Raniwala}}, \bibinfo {author}
  {\bibfnamefont {C.}~\bibnamefont {Gerlach}}, \bibinfo {author} {\bibfnamefont
  {M.}~\bibnamefont {Colangelo}},\ and\ \bibinfo {author} {\bibfnamefont
  {D.}~\bibnamefont {Englund}},\ }\href
  {https://doi.org/10.48550/arXiv.2302.10230} {\bibinfo {title}
  {Cavity-enhanced single artificial atoms in silicon}} (\bibinfo {year}
  {2023}),\ \bibinfo {note} {arXiv:2302.10230 [physics,
  physics:quant-ph]}\BibitemShut {NoStop}%
\bibitem [{\citenamefont {Redjem}\ \emph {et~al.}(2023)\citenamefont {Redjem},
  \citenamefont {Zhiyenbayev}, \citenamefont {Qarony}, \citenamefont {Ivanov},
  \citenamefont {Papapanos}, \citenamefont {Liu}, \citenamefont {Jhuria},
  \citenamefont {Al~Balushi}, \citenamefont {Dhuey}, \citenamefont
  {Schwartzberg}, \citenamefont {Tan}, \citenamefont {Schenkel},\ and\
  \citenamefont {Kanté}}]{redjem_all-silicon_2023}%
  \BibitemOpen
  \bibfield  {author} {\bibinfo {author} {\bibfnamefont {W.}~\bibnamefont
  {Redjem}}, \bibinfo {author} {\bibfnamefont {Y.}~\bibnamefont {Zhiyenbayev}},
  \bibinfo {author} {\bibfnamefont {W.}~\bibnamefont {Qarony}}, \bibinfo
  {author} {\bibfnamefont {V.}~\bibnamefont {Ivanov}}, \bibinfo {author}
  {\bibfnamefont {C.}~\bibnamefont {Papapanos}}, \bibinfo {author}
  {\bibfnamefont {W.}~\bibnamefont {Liu}}, \bibinfo {author} {\bibfnamefont
  {K.}~\bibnamefont {Jhuria}}, \bibinfo {author} {\bibfnamefont {Z.~Y.}\
  \bibnamefont {Al~Balushi}}, \bibinfo {author} {\bibfnamefont
  {S.}~\bibnamefont {Dhuey}}, \bibinfo {author} {\bibfnamefont
  {A.}~\bibnamefont {Schwartzberg}}, \bibinfo {author} {\bibfnamefont {L.~Z.}\
  \bibnamefont {Tan}}, \bibinfo {author} {\bibfnamefont {T.}~\bibnamefont
  {Schenkel}},\ and\ \bibinfo {author} {\bibfnamefont {B.}~\bibnamefont
  {Kanté}},\ }\href {https://doi.org/10.1038/s41467-023-38559-6} {\bibfield
  {journal} {\bibinfo  {journal} {Nature Communications}\ }\textbf {\bibinfo
  {volume} {14}},\ \bibinfo {pages} {3321} (\bibinfo {year}
  {2023})}\BibitemShut {NoStop}%
\bibitem [{\citenamefont {Islam}\ \emph {et~al.}(2024)\citenamefont {Islam},
  \citenamefont {Lee}, \citenamefont {Harper}, \citenamefont {Rahaman},
  \citenamefont {Zhao}, \citenamefont {Vij},\ and\ \citenamefont
  {Waks}}]{islam_cavity-enhanced_2024}%
  \BibitemOpen
  \bibfield  {author} {\bibinfo {author} {\bibfnamefont {F.}~\bibnamefont
  {Islam}}, \bibinfo {author} {\bibfnamefont {C.-M.}\ \bibnamefont {Lee}},
  \bibinfo {author} {\bibfnamefont {S.}~\bibnamefont {Harper}}, \bibinfo
  {author} {\bibfnamefont {M.~H.}\ \bibnamefont {Rahaman}}, \bibinfo {author}
  {\bibfnamefont {Y.}~\bibnamefont {Zhao}}, \bibinfo {author} {\bibfnamefont
  {N.~K.}\ \bibnamefont {Vij}},\ and\ \bibinfo {author} {\bibfnamefont
  {E.}~\bibnamefont {Waks}},\ }\href
  {https://doi.org/10.1021/acs.nanolett.3c04056} {\bibfield  {journal}
  {\bibinfo  {journal} {Nano Letters}\ }\textbf {\bibinfo {volume} {24}},\
  \bibinfo {pages} {319} (\bibinfo {year} {2024})}\BibitemShut {NoStop}%
\bibitem [{\citenamefont {Johnston}\ \emph {et~al.}(2023)\citenamefont
  {Johnston}, \citenamefont {Felix-Rendon}, \citenamefont {Wong},\ and\
  \citenamefont {Chen}}]{johnston_cavity-coupled_2023}%
  \BibitemOpen
  \bibfield  {author} {\bibinfo {author} {\bibfnamefont {A.}~\bibnamefont
  {Johnston}}, \bibinfo {author} {\bibfnamefont {U.}~\bibnamefont
  {Felix-Rendon}}, \bibinfo {author} {\bibfnamefont {Y.-E.}\ \bibnamefont
  {Wong}},\ and\ \bibinfo {author} {\bibfnamefont {S.}~\bibnamefont {Chen}},\
  }\href {https://doi.org/10.48550/arXiv.2310.20014} {\bibinfo {title}
  {Cavity-coupled telecom atomic source in silicon}} (\bibinfo {year} {2023}),\
  \bibinfo {note} {arXiv:2310.20014 [physics, physics:quant-ph]}\BibitemShut
  {NoStop}%
\bibitem [{\citenamefont {Lee}\ \emph {et~al.}(2023)\citenamefont {Lee},
  \citenamefont {Islam}, \citenamefont {Harper}, \citenamefont {Buyukkaya},
  \citenamefont {Higginbottom}, \citenamefont {Simmons},\ and\ \citenamefont
  {Waks}}]{lee_high-efficiency_2023}%
  \BibitemOpen
  \bibfield  {author} {\bibinfo {author} {\bibfnamefont {C.-M.}\ \bibnamefont
  {Lee}}, \bibinfo {author} {\bibfnamefont {F.}~\bibnamefont {Islam}}, \bibinfo
  {author} {\bibfnamefont {S.}~\bibnamefont {Harper}}, \bibinfo {author}
  {\bibfnamefont {M.~A.}\ \bibnamefont {Buyukkaya}}, \bibinfo {author}
  {\bibfnamefont {D.}~\bibnamefont {Higginbottom}}, \bibinfo {author}
  {\bibfnamefont {S.}~\bibnamefont {Simmons}},\ and\ \bibinfo {author}
  {\bibfnamefont {E.}~\bibnamefont {Waks}},\ }\href
  {https://doi.org/10.1021/acsphotonics.3c01142} {\bibfield  {journal}
  {\bibinfo  {journal} {ACS Photonics}\ }\textbf {\bibinfo {volume} {10}},\
  \bibinfo {pages} {3844} (\bibinfo {year} {2023})}\BibitemShut {NoStop}%
\bibitem [{\citenamefont {Thonke}\ \emph {et~al.}(1981)\citenamefont {Thonke},
  \citenamefont {Klemisch}, \citenamefont {Weber},\ and\ \citenamefont
  {Sauer}}]{thonke_new_1981}%
  \BibitemOpen
  \bibfield  {author} {\bibinfo {author} {\bibfnamefont {K.}~\bibnamefont
  {Thonke}}, \bibinfo {author} {\bibfnamefont {H.}~\bibnamefont {Klemisch}},
  \bibinfo {author} {\bibfnamefont {J.}~\bibnamefont {Weber}},\ and\ \bibinfo
  {author} {\bibfnamefont {R.}~\bibnamefont {Sauer}},\ }\href
  {https://doi.org/10.1103/PhysRevB.24.5874} {\bibfield  {journal} {\bibinfo
  {journal} {Physical Review B}\ }\textbf {\bibinfo {volume} {24}},\ \bibinfo
  {pages} {5874} (\bibinfo {year} {1981})}\BibitemShut {NoStop}%
\bibitem [{\citenamefont {Davies}\ \emph {et~al.}(1983)\citenamefont {Davies},
  \citenamefont {Lightowlers},\ and\ \citenamefont
  {Carmo}}]{davies_carbon-related_1983}%
  \BibitemOpen
  \bibfield  {author} {\bibinfo {author} {\bibfnamefont {G.}~\bibnamefont
  {Davies}}, \bibinfo {author} {\bibfnamefont {E.~C.}\ \bibnamefont
  {Lightowlers}},\ and\ \bibinfo {author} {\bibfnamefont {M.~d.}\ \bibnamefont
  {Carmo}},\ }\href {https://doi.org/10.1088/0022-3719/16/28/017} {\bibfield
  {journal} {\bibinfo  {journal} {Journal of Physics C: Solid State Physics}\
  }\textbf {\bibinfo {volume} {16}},\ \bibinfo {pages} {5503} (\bibinfo {year}
  {1983})}\BibitemShut {NoStop}%
\bibitem [{\citenamefont {Davies}(1989)}]{davies_optical_1989}%
  \BibitemOpen
  \bibfield  {author} {\bibinfo {author} {\bibfnamefont {G.}~\bibnamefont
  {Davies}},\ }\href {https://doi.org/10.1016/0370-1573(89)90064-1} {\bibfield
  {journal} {\bibinfo  {journal} {Physics Reports}\ }\textbf {\bibinfo {volume}
  {176}},\ \bibinfo {pages} {83} (\bibinfo {year} {1989})}\BibitemShut
  {NoStop}%
\bibitem [{\citenamefont {Beaufils}\ \emph {et~al.}(2018)\citenamefont
  {Beaufils}, \citenamefont {Redjem}, \citenamefont {Rousseau}, \citenamefont
  {Jacques}, \citenamefont {Kuznetsov}, \citenamefont {Raynaud}, \citenamefont
  {Voisin}, \citenamefont {Benali}, \citenamefont {Herzig}, \citenamefont
  {Pezzagna}, \citenamefont {Meijer}, \citenamefont {Abbarchi},\ and\
  \citenamefont {Cassabois}}]{beaufils_optical_2018}%
  \BibitemOpen
  \bibfield  {author} {\bibinfo {author} {\bibfnamefont {C.}~\bibnamefont
  {Beaufils}}, \bibinfo {author} {\bibfnamefont {W.}~\bibnamefont {Redjem}},
  \bibinfo {author} {\bibfnamefont {E.}~\bibnamefont {Rousseau}}, \bibinfo
  {author} {\bibfnamefont {V.}~\bibnamefont {Jacques}}, \bibinfo {author}
  {\bibfnamefont {A.~Y.}\ \bibnamefont {Kuznetsov}}, \bibinfo {author}
  {\bibfnamefont {C.}~\bibnamefont {Raynaud}}, \bibinfo {author} {\bibfnamefont
  {C.}~\bibnamefont {Voisin}}, \bibinfo {author} {\bibfnamefont
  {A.}~\bibnamefont {Benali}}, \bibinfo {author} {\bibfnamefont
  {T.}~\bibnamefont {Herzig}}, \bibinfo {author} {\bibfnamefont
  {S.}~\bibnamefont {Pezzagna}}, \bibinfo {author} {\bibfnamefont
  {J.}~\bibnamefont {Meijer}}, \bibinfo {author} {\bibfnamefont
  {M.}~\bibnamefont {Abbarchi}},\ and\ \bibinfo {author} {\bibfnamefont
  {G.}~\bibnamefont {Cassabois}},\ }\href
  {https://doi.org/10.1103/PhysRevB.97.035303} {\bibfield  {journal} {\bibinfo
  {journal} {Physical Review B}\ }\textbf {\bibinfo {volume} {97}},\ \bibinfo
  {pages} {035303} (\bibinfo {year} {2018})}\BibitemShut {NoStop}%
\bibitem [{\citenamefont {Lefaucher}\ \emph {et~al.}(2023)\citenamefont
  {Lefaucher}, \citenamefont {Jager}, \citenamefont {Calvo}, \citenamefont
  {Durand}, \citenamefont {Baron}, \citenamefont {Cache}, \citenamefont
  {Jacques}, \citenamefont {Robert-Philip}, \citenamefont {Cassabois},
  \citenamefont {Herzig}, \citenamefont {Meijer}, \citenamefont {Pezzagna},
  \citenamefont {Khoury}, \citenamefont {Abbarchi}, \citenamefont {Dréau},\
  and\ \citenamefont {Gérard}}]{lefaucher_cavity-enhanced_2023}%
  \BibitemOpen
  \bibfield  {author} {\bibinfo {author} {\bibfnamefont {B.}~\bibnamefont
  {Lefaucher}}, \bibinfo {author} {\bibfnamefont {J.-B.}\ \bibnamefont
  {Jager}}, \bibinfo {author} {\bibfnamefont {V.}~\bibnamefont {Calvo}},
  \bibinfo {author} {\bibfnamefont {A.}~\bibnamefont {Durand}}, \bibinfo
  {author} {\bibfnamefont {Y.}~\bibnamefont {Baron}}, \bibinfo {author}
  {\bibfnamefont {F.}~\bibnamefont {Cache}}, \bibinfo {author} {\bibfnamefont
  {V.}~\bibnamefont {Jacques}}, \bibinfo {author} {\bibfnamefont
  {I.}~\bibnamefont {Robert-Philip}}, \bibinfo {author} {\bibfnamefont
  {G.}~\bibnamefont {Cassabois}}, \bibinfo {author} {\bibfnamefont
  {T.}~\bibnamefont {Herzig}}, \bibinfo {author} {\bibfnamefont
  {J.}~\bibnamefont {Meijer}}, \bibinfo {author} {\bibfnamefont
  {S.}~\bibnamefont {Pezzagna}}, \bibinfo {author} {\bibfnamefont
  {M.}~\bibnamefont {Khoury}}, \bibinfo {author} {\bibfnamefont
  {M.}~\bibnamefont {Abbarchi}}, \bibinfo {author} {\bibfnamefont
  {A.}~\bibnamefont {Dréau}},\ and\ \bibinfo {author} {\bibfnamefont {J.-M.}\
  \bibnamefont {Gérard}},\ }\href {https://doi.org/10.1063/5.0130196}
  {\bibfield  {journal} {\bibinfo  {journal} {Applied Physics Letters}\
  }\textbf {\bibinfo {volume} {122}},\ \bibinfo {pages} {061109} (\bibinfo
  {year} {2023})}\BibitemShut {NoStop}%
\bibitem [{\citenamefont {Beveratos}\ \emph {et~al.}(2002)\citenamefont
  {Beveratos}, \citenamefont {Kühn}, \citenamefont {Brouri}, \citenamefont
  {Gacoin}, \citenamefont {Poizat},\ and\ \citenamefont
  {Grangier}}]{beveratos_room_2002}%
  \BibitemOpen
  \bibfield  {author} {\bibinfo {author} {\bibfnamefont {A.}~\bibnamefont
  {Beveratos}}, \bibinfo {author} {\bibfnamefont {S.}~\bibnamefont {Kühn}},
  \bibinfo {author} {\bibfnamefont {R.}~\bibnamefont {Brouri}}, \bibinfo
  {author} {\bibfnamefont {T.}~\bibnamefont {Gacoin}}, \bibinfo {author}
  {\bibfnamefont {J.-P.}\ \bibnamefont {Poizat}},\ and\ \bibinfo {author}
  {\bibfnamefont {P.}~\bibnamefont {Grangier}},\ }\href
  {https://doi.org/10.1140/epjd/e20020023} {\bibfield  {journal} {\bibinfo
  {journal} {The European Physical Journal D}\ }\textbf {\bibinfo {volume}
  {18}},\ \bibinfo {pages} {191} (\bibinfo {year} {2002})}\BibitemShut
  {NoStop}%
\bibitem [{\citenamefont {Durand}\ \emph {et~al.}(2024)\citenamefont {Durand},
  \citenamefont {Baron},\ and\ \citenamefont {et~al.
  In~preparation.}}]{durand_notitle_2024}%
  \BibitemOpen
  \bibfield  {author} {\bibinfo {author} {\bibfnamefont {A.}~\bibnamefont
  {Durand}}, \bibinfo {author} {\bibfnamefont {Y.}~\bibnamefont {Baron}},\ and\
  \bibinfo {author} {\bibnamefont {et~al. In~preparation.}},\ }\href@noop {} {}
  (\bibinfo {year} {2024})\BibitemShut {NoStop}%
\bibitem [{\citenamefont {Foy}\ \emph {et~al.}(1981)\citenamefont {Foy},
  \citenamefont {Carmo}, \citenamefont {Davies},\ and\ \citenamefont
  {Lightowlers}}]{foy_uniaxial_1981}%
  \BibitemOpen
  \bibfield  {author} {\bibinfo {author} {\bibfnamefont {C.~P.}\ \bibnamefont
  {Foy}}, \bibinfo {author} {\bibfnamefont {M.~C.~d.}\ \bibnamefont {Carmo}},
  \bibinfo {author} {\bibfnamefont {G.}~\bibnamefont {Davies}},\ and\ \bibinfo
  {author} {\bibfnamefont {E.~C.}\ \bibnamefont {Lightowlers}},\ }\href
  {https://doi.org/10.1088/0022-3719/14/1/002} {\bibfield  {journal} {\bibinfo
  {journal} {Journal of Physics C: Solid State Physics}\ }\textbf {\bibinfo
  {volume} {14}},\ \bibinfo {pages} {L7} (\bibinfo {year} {1981})},\ \bibinfo
  {note} {publisher: IOP Publishing}\BibitemShut {NoStop}%
\bibitem [{\citenamefont {O'Donnell}\ \emph {et~al.}(1983)\citenamefont
  {O'Donnell}, \citenamefont {Lee},\ and\ \citenamefont
  {Watkins}}]{odonnell_origin_1983}%
  \BibitemOpen
  \bibfield  {author} {\bibinfo {author} {\bibfnamefont {K.~P.}\ \bibnamefont
  {O'Donnell}}, \bibinfo {author} {\bibfnamefont {K.~M.}\ \bibnamefont {Lee}},\
  and\ \bibinfo {author} {\bibfnamefont {G.~D.}\ \bibnamefont {Watkins}},\
  }\href {https://doi.org/10.1016/0378-4363(83)90256-5} {\bibfield  {journal}
  {\bibinfo  {journal} {Physica B+C}\ }\textbf {\bibinfo {volume} {116}},\
  \bibinfo {pages} {258} (\bibinfo {year} {1983})}\BibitemShut {NoStop}%
\bibitem [{\citenamefont {Song}\ \emph {et~al.}(1990)\citenamefont {Song},
  \citenamefont {Zhan}, \citenamefont {Benson},\ and\ \citenamefont
  {Watkins}}]{song_bistable_1990}%
  \BibitemOpen
  \bibfield  {author} {\bibinfo {author} {\bibfnamefont {L.~W.}\ \bibnamefont
  {Song}}, \bibinfo {author} {\bibfnamefont {X.~D.}\ \bibnamefont {Zhan}},
  \bibinfo {author} {\bibfnamefont {B.~W.}\ \bibnamefont {Benson}},\ and\
  \bibinfo {author} {\bibfnamefont {G.~D.}\ \bibnamefont {Watkins}},\ }\href
  {https://doi.org/10.1103/PhysRevB.42.5765} {\bibfield  {journal} {\bibinfo
  {journal} {Physical Review B}\ }\textbf {\bibinfo {volume} {42}},\ \bibinfo
  {pages} {5765} (\bibinfo {year} {1990})}\BibitemShut {NoStop}%
\bibitem [{\citenamefont {Udvarhelyi}\ \emph {et~al.}(2021)\citenamefont
  {Udvarhelyi}, \citenamefont {Somogyi}, \citenamefont {Thiering},\ and\
  \citenamefont {Gali}}]{udvarhelyi_identification_2021}%
  \BibitemOpen
  \bibfield  {author} {\bibinfo {author} {\bibfnamefont {P.}~\bibnamefont
  {Udvarhelyi}}, \bibinfo {author} {\bibfnamefont {B.}~\bibnamefont {Somogyi}},
  \bibinfo {author} {\bibfnamefont {G.}~\bibnamefont {Thiering}},\ and\
  \bibinfo {author} {\bibfnamefont {A.}~\bibnamefont {Gali}},\ }\href
  {https://doi.org/10.1103/PhysRevLett.127.196402} {\bibfield  {journal}
  {\bibinfo  {journal} {Physical Review Letters}\ }\textbf {\bibinfo {volume}
  {127}},\ \bibinfo {pages} {196402} (\bibinfo {year} {2021})}\BibitemShut
  {NoStop}%
\bibitem [{\citenamefont {Lee}\ \emph {et~al.}(1982)\citenamefont {Lee},
  \citenamefont {O'Donnell}, \citenamefont {Weber}, \citenamefont {Cavenett},\
  and\ \citenamefont {Watkins}}]{lee_optical_1982}%
  \BibitemOpen
  \bibfield  {author} {\bibinfo {author} {\bibfnamefont {K.~M.}\ \bibnamefont
  {Lee}}, \bibinfo {author} {\bibfnamefont {K.~P.}\ \bibnamefont {O'Donnell}},
  \bibinfo {author} {\bibfnamefont {J.}~\bibnamefont {Weber}}, \bibinfo
  {author} {\bibfnamefont {B.~C.}\ \bibnamefont {Cavenett}},\ and\ \bibinfo
  {author} {\bibfnamefont {G.~D.}\ \bibnamefont {Watkins}},\ }\href
  {https://doi.org/10.1103/PhysRevLett.48.37} {\bibfield  {journal} {\bibinfo
  {journal} {Physical Review Letters}\ }\textbf {\bibinfo {volume} {48}},\
  \bibinfo {pages} {37} (\bibinfo {year} {1982})}\BibitemShut {NoStop}%
\end{thebibliography}%

\end{document}